\documentclass[aip,cha,reprint,superscriptaddress,floatfix]{revtex4-2}
\usepackage{graphicx}
\usepackage{amsmath,amssymb,mathtools}
\usepackage[T1]{fontenc}
\usepackage[utf8]{inputenc}
\usepackage{color}

\begin{document}

\title{Directed Wada cascade in systems with discrete scale invariance}

\author{P. Haerter}
\email{haerter@usp.br}
\affiliation{Instituto de Física, Universidade de São Paulo, São Paulo, São Paulo, 05508-090, Brazil}

\author{R. L. Viana}%
\affiliation{Departamento de Física, Universidade Federal do Paraná, Curitiba, 81531-990, Paraná, Brazil.}
\affiliation{Centro Interdisciplinar de Ciência, Tecnologia e Inovação, Núcleo de Modelagem e Computação Científica, Universidade Federal do Paraná, Curitiba, 81531-990, Paraná, Brazil}

\author{M. A.F. Sanjuán}
\affiliation{Nonlinear Dynamics, Chaos and Complex Systems Group, Departamento de F\'{i}sica, Universidad Rey Juan Carlos, Tulip\'{a}n s/n, 28933 M\'{o}stoles, Madrid, Spain}
\affiliation{Royal Academy of Sciences of Spain, 28004 Madrid, Spain}

\author{I. L. Caldas}
\affiliation{Instituto de Física, Universidade de São Paulo, São Paulo, São Paulo, 05508-090, Brazil}

\begin{abstract}
Dissipative systems whose equations are unchanged by a dilation can host an infinite ladder of attractors, each shell an exact magnified copy of the one before. Since such dynamics cannot tell a magnification from its inverse, the natural expectation is that these basins are maximally entangled, with every boundary point lying on the boundary of all of them, following the Wada property. We find that this expectation fails, and fails in a precise way. Dilation covariance and a single trapping region together bound which basin boundaries a point can lie on, and the bound depends only on its distance to the origin. No point lies on the boundary of all of them. The basins on whose boundary a point does lie are the deeper ones, as we prove for the stable manifolds of the boundary saddles and support numerically for the rest of the boundary, so the boundaries nest in one direction only. We call this structure a directed Wada cascade, and the standard Wada tests, as usually applied, do not detect it. It arises in maps whose nonlinearity is a log-periodic modulation of a homogeneous function, because the same periodicity builds the ladder through mode locking and confines the orbits. We show the cascade in maps that differ in invertibility, nonlinearity, scaling exponent and dimension, and in a flow.
\end{abstract}

\maketitle

\begin{quotation}
A dissipative system whose equations look the same after a magnification by a fixed factor and its powers, a symmetry called discrete scale invariance, can hold an endless ladder of attractors, like nested matryoshka dolls. Every initial condition ends on one of them, and which one depends on how the regions that lead to each attractor, the basins, are arranged. The symmetry suggests the most entangled arrangement, the Wada property, and the standard tests agree, yet we show that this is not so. The basins are nested in one direction only, toward the origin, a structure we call a directed Wada cascade. We find it in maps, in a flow and in a high-dimensional network, and it makes the uncertainty about the final state unbounded toward the origin but bounded outward.
\end{quotation}

\section{Introduction}
\label{sec:intro}

Some systems look the same when magnified, and some only when magnified by a fixed factor $S$ and its powers. This discrete scale invariance (DSI) \cite{sornetteDiscretescaleInvarianceComplex1998} can be a symmetry of the equations of motion of a dissipative system. A magnification then maps an attractor onto an attractor, and the system holds a ladder of attractors $A_k=S^{-k}A_0$, $k\in\mathbb{Z}$, whose members we call shells. This differs from the more familiar case of a discrete translation, which carries one attractor onto the next, makes the basins copies of one another, and passes the symmetry to the basin boundaries in full \cite{sprott2017megastability,wangInfiniteNumberWada2023,xieSymmetryGroupsBasins2025}. A magnification has a fixed point, the origin, and the shells pile up toward it.

The basin boundaries decide which shell an initial condition reaches. Basins of attraction with fractal properties have been investigated in many practical applications, for example in mechanical oscillations with limited power supply \cite{desouzaBasinsAttractionChanges2005}, wind-induced oscillations of suspension bridges \cite{defreitasMultistabilityBasinBoundary2004}, and rattling in gearboxes \cite{desouzaNoiseinducedBasinHopping2005}. The strongest form of final-state uncertainty is the Wada property, in which every boundary point of one basin is a boundary point of all the others \cite{kennedyBasinsWada1991,nusseWadaBasinBoundaries1996,aguirreFractalStructuresNonlinear2009,dazaMultistabilityUnpredictability2024}. For a ladder of shells the symmetry points to it, because the dynamics treats $S$ and $S^{-1}$ alike and no shell is preferred. One would therefore expect every boundary of the ladder to be Wada.

DSI has been identified in Efimov states \cite{braatenUniversalityFewbodySystems2006,kraemerEvidenceEfimovQuantum2006,zaccantiObservationEfimovSpectrum2009}, in the magnetoresistance of topological semimetals \cite{wangDiscoveryLogperiodicOscillations2018,wangLogperiodicQuantumMagnetooscillations2019}, in turbulent dissipation \cite{zhouNEWEVIDENCEDISCRETE2003}, before earthquakes \cite{johansenNewEvidenceEarthquake2000} and before material failure \cite{anifrani1995universal}. These are spectral and scaling signatures, and which attractor an initial condition reaches has not been asked in any of them. In nonlinear dynamics, exactly self-similar nested attractors are known in flows
\cite{karimov2024matryoshka,karimovExtremelyMultistableMatryoshka2025,wangFractalInducedMultistabilityChuas2025} whose right-hand side is discrete scale invariant, a phenomenon named matryoshka multistability \cite{karimov2024matryoshka,karimovExtremelyMultistableMatryoshka2025,wangFractalInducedMultistabilityChuas2025,burkinMatryoshkaMultistableSystem2026}, and earlier in flows with nested scroll attractors \cite{escalanteGonzalezMultistableSystemsHidden2020,escalanteGonzalezMultistableSystemsNested2022}. These works build the ladder in continuous time. We found no map with an exact dilation symmetry and such a ladder in the literature, and which basin boundaries contain a given boundary point of such a ladder is open.

We show that the boundaries of the ladder are not Wada. The magnification symmetry, together with a single trapping region, bounds which basin boundaries a point can lie on, and the bound depends only on the distance of the point to the origin. No boundary point lies on the boundary of all the basins, so the boundary is non-Wada at every point. The basins on whose boundary a point does lie are the deeper ones, as we prove for the stable manifolds of the boundary saddles and assume, with numerical support, for the rest of the boundary. The dynamics treats $S$ and $S^{-1}$ alike, yet the nesting of the boundaries runs in one direction only. In contrast with a discrete translation, the boundaries therefore inherit only half of the symmetry. We call this structure a directed Wada cascade. Maps whose nonlinearity is a log-periodic modulation of a homogeneous function generate it, because the same modulation builds the ladder through mode locking and also confines the orbits.

We do not claim a new topological type of boundary, since partially Wada boundaries are known \cite{zhangWadaBifurcationsPartially2013}. What is new is the mechanism behind the nesting, the class of maps that generates it, and the fact that the usual diagnostics \cite{dazaTestingBasinsWada2015,dazaAscertainingWhenBasin2018,HowTodetecWadabasins} report Wada where the structure is not.

Section~\ref{sec:glance} shows the whole picture at a glance, in plain terms and with the main results on display. The remaining sections make it precise. Section~\ref{sec:mech} derives the mechanism on one map. It defines what covariance and confinement do to the basins, the exact limit they impose, how invariant manifolds make the nesting run one way, and what the usual diagnostics see. Section~\ref{sec:class} presents the class of maps that has the mechanism, showing how a log-periodic nonlinearity yields at once a ladder of attractors and a ladder of trapping regions. Section~\ref{sec:examples} gives the examples, two more planar maps that differ from the first in as many respects as we could arrange, a flow of three variables, and a network of higher dimension. Finally, Section~\ref{sec:disc} states what is proved, what is numerical and what is open.

\section{The cascade at a glance}
\label{sec:glance}

Think of a set of matryoshka dolls, each one an exact copy of the next, smaller by the same factor, an image already used for flows with fractal nonlinearities \cite{karimov2024matryoshka}. A dynamical system has this structure when its equations do not change if the state is magnified by a fixed factor $S$. If $\mathbf{x}\to F(\mathbf{x})$ is the rule that advances the system by one step, this means $F(S\mathbf{x})=S\,F(\mathbf{x})$. An orbit magnified by $S$ is then still an orbit, so an attractor magnified by $S$ is still an attractor, and the ladder of shells $A_k=S^{-k}A_0$ of the Introduction follows. We call $B_k$ the basin of $A_k$, the set of initial conditions that end on it, and $\partial B_k$ its boundary. Both scale as the attractor does, $B_k=S^{-k}B_0$ and $\partial B_k=S^{-k}\partial B_0$. This is not the self-similarity of a single attractor, as in the Hénon map. There, zooming into one attractor shows the same structure again, only approximately and with no fixed ratio. Here the symmetry belongs to the equations, it relates different attractors, and the ratio $S$ is exact and applies only to its integer powers.

The simplest example we found is a map of the plane, which we call the matryoshka map,
\begin{subequations}\label{eq:map}
\begin{align}
x_{n+1}&=y_n,\\
y_{n+1}&=-b\,x_n+y_n\,\bigl[\alpha+\beta\cos(\omega\ln|y_n|)\bigr],
\end{align}
\end{subequations}
with $\alpha=-0.5$, $\beta=1.08$, $b=0.1$, $\omega=\pi$, and the second term of $y_{n+1}$ set to zero at $y_n=0$. The factor that multiplies $y_n$ is a constant plus an oscillation in the logarithm of $|y_n|$, a log-periodic function. A magnification by $S=e^{2\pi/\omega}=e^{2}\simeq7.4$ shifts $\ln|y_n|$ by exactly one period, so the factor does not notice it and $F(S\mathbf{x})=S\,F(\mathbf{x})$ holds. The origin is the only fixed point, yet generic orbits do not collapse onto it. They settle on one of the chaotic attractors of the ladder, one in each shell, and each of them is hidden, because its basin stays away from the neighbourhood of the origin.

\begin{figure}[t]
    \centering
    \includegraphics[width=\columnwidth]{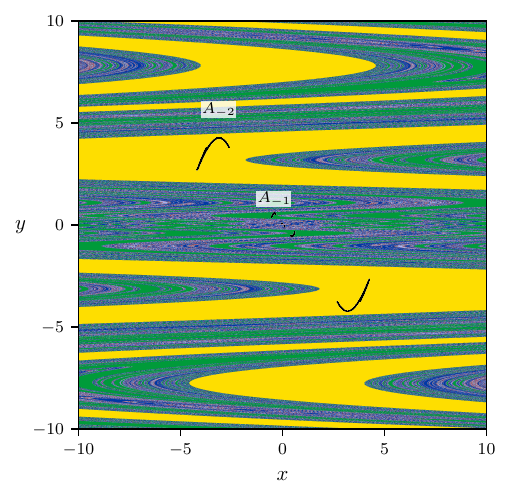}
    \caption{\textbf{Basins of attraction of the matryoshka map on $[-10,10]^2$.} Colours label the shell $k$ of the attractor reached, as in every figure of the paper. The black curves are the attractors $A_{-2}$ and $A_{-1}$, and the smaller ones lie too close to the origin to be seen. The picture repeats around the origin, shrunk by $S=e^{2}$ at each shell.}
    \label{fig:glance}
\end{figure}

\begin{figure*}[t]
    \centering
    \includegraphics[width=\textwidth]{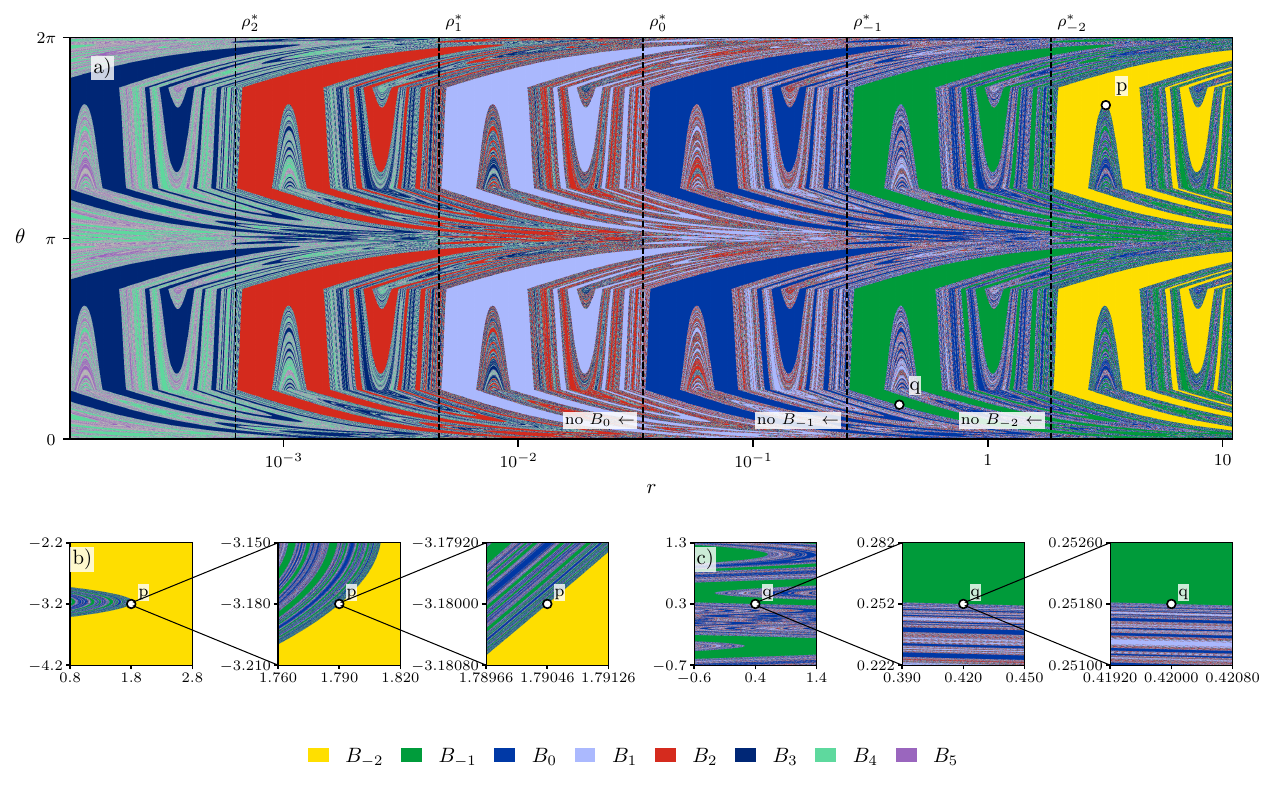}
    \caption{\textbf{The directed Wada cascade in the matryoshka map.} Colours label the shell index $k$ of the attractor reached and are the same in every figure. (a) Basins against the distance $r$ to the origin, on a logarithmic axis, and the polar angle $\theta$. In this coordinate system a dilation is a translation by $\log_{10}S\simeq0.87$, the pattern repeats, and the basin $B_k$ has a vertical edge at $r=\rho^{*}_k$ (dashed lines), with no $B_k$ to its left. The circles $p$ and $q$ are boundary points at distances $3.18$ and $0.42$. (b,c) Magnifications around $p$ (b) and $q$ (c). At $p$ the outermost basin $B_{-2}$ fills one smooth side and all deeper basins mix on the other, at every width. At $q$ the outermost basin is $B_{-1}$, and $B_{-2}$ never appears.}
    \label{fig:cascade}
\end{figure*}

Figure~\ref{fig:glance} paints each initial condition with the colour of the shell it reaches, and the black curves are the attractors. The basins are not compact blobs. They are drawn into thin filaments that interleave at every magnification, being fractal basins \cite{mcdonaldFractalBasinBoundaries1985}. What is special here is that the picture repeats. Around the origin the same pattern appears again, shrunk by $S$ each time, and each repetition brings a new attractor with its own basin. A logarithmic view makes this explicit. If the basins are plotted against the logarithm of the distance $r$ to the origin and the polar angle $\theta$, a magnification by $S$ becomes a plain shift, by $\log_{10}S\simeq0.87$ in Fig.~\ref{fig:cascade}a, and the picture is periodic from side to side with the colours stepping by one shell per period.

A point on a smooth basin boundary lies on the boundary of exactly two basins. The Wada property is the opposite extreme, in which every point of the boundary of one basin is also a point of the boundary of all the others. However closely one looks at such a point, every basin is still there. A ladder of attractors that treats a magnification and its inverse alike, with no preferred shell, is a natural place to expect this property, and the standard tests agree. The first test counts how many basins are found in a small disc around a boundary point. In the matryoshka map, more than $98\%$ of the boundary points have three or more basins within a few percent of their own distance $r$ to the origin, at every $r$. The second, the merging test, relabels two basins as one and measures how much of the boundary disappears. For a Wada boundary almost nothing is lost, and in the matryoshka map less than $0.04\%$ is lost, for any pair \cite{dazaTestingBasinsWada2015,dazaAscertainingWhenBasin2018}. By these measures the matryoshka map is a Wada system.

The tests look at one scale at a time. The picture changes when we follow particular boundary points all the way down in scale (Fig.~\ref{fig:cascade}b,c). The point $p$ is at distance $3.18$ from the origin. Zooming in on it, the outermost basin $B_{-2}$ (yellow) fills one smooth side of the boundary and all the deeper basins mix on the other side, however far we zoom. The point $q$ is at distance $0.42$. It shows the same mixing, but its outermost basin is $B_{-1}$ and the yellow basin never appears, at any magnification. A Wada boundary would show every basin at both points. A boundary point close to the origin therefore cannot lie on the boundary of the basins of the largest attractors, and the closer it is the fewer boundaries it lies on. Yet it lies on the boundary of all the smaller ones, infinitely many, as proved for the stable manifolds and supported numerically for the rest of the boundary. The tests miss this because there are always at least three basins nearby.

We call this structure a directed Wada cascade. It has three properties, stated precisely in Sec.~\ref{sec:mech}.
\begin{enumerate}
\item[(i)] Nesting. A point on the boundary of the basin of one shell lies on the boundary of the basins of all deeper shells.
\item[(ii)] Finite reach. The outermost basin on whose boundary a point can lie is set by its distance to the origin, so no point lies on the boundary of all basins.
\item[(iii)] Strictness. The boundaries of consecutive basins differ, $\partial B_k\neq\partial B_{k+1}$, and with (i) the deeper one has points that the outer one lacks.
\end{enumerate}
Demagnifying a boundary by $S$ gives the boundary of the next deeper basin, which contains the original, while magnifying it returns only a part. The structure is Wada in one direction only, toward the origin, and this is why we call it directed.

\begin{figure}[t]
    \centering
    \includegraphics[width=0.9\columnwidth]{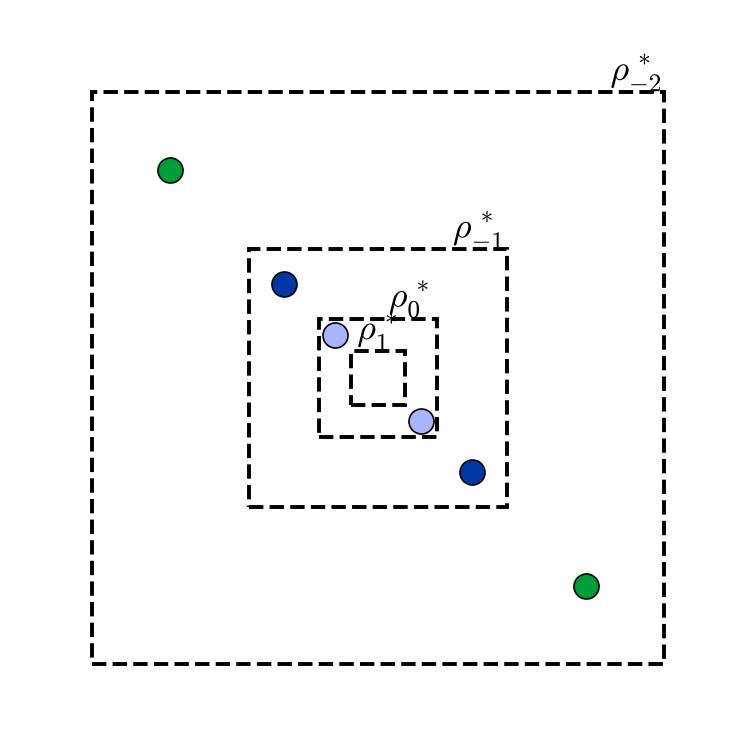}
    \caption{\textbf{Sketch of the ladder of trapping squares in the matryoshka map.} The dashed squares have half-widths $\rho^{*}_k$ and each one is trapping. A point that starts inside the square $\rho^{*}_k$ stays there, so it can only reach the attractors drawn inside it, never the attractor of shell $k$ or of any outer shell. Dots mark the attractor of each shell, with the colours of Fig.~\ref{fig:cascade}.}
    \label{fig:squares}
\end{figure}

Two ingredients are enough for this to happen. The first is the symmetry, and the second is a trapping square, a region of phase space that orbits can enter but never leave. The matryoshka map has such squares around the origin, and since the symmetry carries a trapping square into another one magnified by $S$, they come as a ladder with one square for each shell. For each shell take the square that is just smaller than its attractor, with half-width $\rho^{*}_k=S^{-k}\rho^{*}_0$ (Fig.~\ref{fig:squares}). An orbit that starts inside it is trapped there forever, so it can reach the smaller attractors inside but never $A_k$ nor any larger one. A boundary point inside the square has a small disc around it that stays inside and holds no point of $B_k$, so the point cannot lie on the boundary of $B_k$. This is the finite reach. A point at distance $r$ from the origin can lie only on the boundaries of the basins with
\begin{equation*}
k\ge\log_S\frac{\rho^{*}_0}{r},
\end{equation*}
where $\rho^{*}_0=0.0341$ for the matryoshka map.

\begin{figure}[t]
    \centering
    \includegraphics[width=\columnwidth]{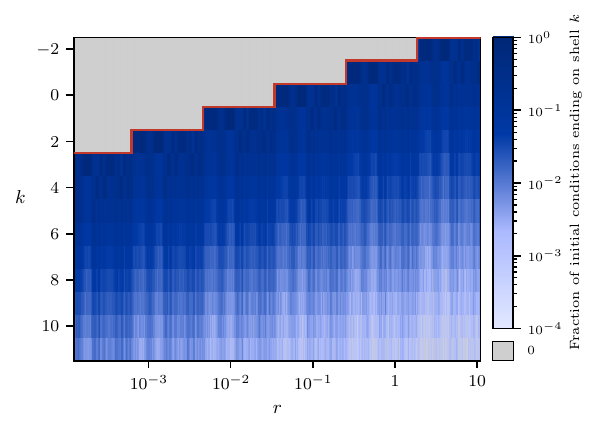}
    \caption{\textbf{Fraction of the initial conditions at distance $r$ that end on the attractor of shell $k$.} The red staircase is the reach law, Eq.~(\ref{eq:reach}), $k\ge\log_S(\rho^{*}_0/r)$ with $k$ an integer. The largest attractor reached steps down by one shell each time $r$ shrinks by $S$, while in a Wada boundary this envelope would be flat. Grey cells are those where no initial condition ended. }
    \label{fig:tests}
\end{figure}

This prediction is easy to test. Prepare the system at distance $r$ from the origin, record the attractor it reaches, and repeat for many initial conditions and many distances. Figure~\ref{fig:tests} shows the outcome. The largest attractor that is reached steps down by one shell each time $r$ shrinks by a factor $S$, exactly along the staircase given by the bound, and the cells above it are empty. In a Wada system this envelope would be flat. The uncertainty is therefore unbounded inward and bounded outward. From a distance $r$ the outcome may be any of infinitely many small attractors, but never one larger than the limit set by $r$.

The nesting has a different origin, in the invariant manifolds of the map. Every shell has a saddle orbit whose stable manifold, the set of points that converge to it, lies in the boundary of the basin of that shell. In the matryoshka map the unstable manifold of the saddle of one shell crosses the stable manifold of the saddle of the next, deeper shell, and this crossing makes the boundary of the outer basin part of the boundary of the deeper one. The symmetry alone would allow the reverse as well, and the trapping squares forbid it.

Nothing above is special to the matryoshka map. The recipe is a linear map plus a nonlinear term whose strength is a bounded function of the logarithm of the amplitude, periodic with period $\ln S$. In compact form, $F(\mathbf{x})=L\mathbf{x}+W\,H(\mathbf{x})\,\mathbf{x}$ with constant matrices $L$ and $W$ and $H(S\mathbf{x})=H(\mathbf{x})$. Covariance is then automatic. The same periodicity builds the ladder of attractors, because the orbit locks onto the oscillation of the gain, and the ladder of trapping squares, because between two peaks of the oscillation the square grows while the largest push of the nonlinear term does not.

 \section{Wada cascade mechanism}
\label{sec:mech}

To derive all the properties of the directed Wada cascade, we start with a map $F$ on $\mathbb{R}^{d}$. It is dilation covariant with ratio $S>1$ if
\begin{equation}
F(S\,\mathbf{x})=S\,F(\mathbf{x}).
\label{eq:cov}
\end{equation}
Every invariant set then has an exact copy magnified by $S$, with the same stability. The Jacobian satisfies $J(S\mathbf{x})=J(\mathbf{x})$, so the multipliers of $S$-related orbits are identical. If $A_0$ is an attractor, so is $A_k=S^{-k}A_0$, with basin $B_k=S^{-k}B_0$ and boundary $\partial B_k=S^{-k}\partial B_0$, where the index $k$ grows toward the origin. We call $B_{k'}$ deeper than $B_k$ if $k'>k$. A point $\mathbf{x}$ lies on the boundary of $B_k$ if $\mathbf{x}\in\partial B_k$, that is, if every neighbourhood of $\mathbf{x}$, however small, contains points of $B_k$. The reach $m(\mathbf{x})$ of $\mathbf{x}$ is the outermost basin whose boundary contains $\mathbf{x}$, the one with the smallest $k$.

A second key notion is that of a trapping region, a set $R$ that the map cannot leave once a trajectory has entered it, $F(R)\subseteq R$. Trapping regions are what confine the dynamics, and the simplest ones are balls $R(\rho)=\{\mathbf{x}:\|\mathbf{x}\|\le\rho\}$ of radius $\rho$, in a norm chosen for each system. The distance $r$ of a point to the origin is measured in the same norm. Discrete scale invariance extends this to a whole family of balls. Since $F$ commutes with the rescaling by $S$, if $R(\rho)$ is trapping then
\begin{equation}
F[R(S\rho)] = S\,F[R(\rho)] \subseteq S\,R(\rho) = R(S\rho),
\end{equation}
so $R(S\rho)$ is trapping too. The set of trapping radii is therefore invariant under $\rho\to S\rho$, and confinement, like the attractors, appears as a ladder of nested regions $R(S^{-k}\rho)$.

All the ideas below are developed on the matryoshka map, Eq.~(\ref{eq:map}), which has scale invariance and the origin as its only fixed point. Its properties are explored in Sec.~\ref{sec:map1}.

The mechanism needs trapping balls, and for the matryoshka map they can be found explicitly. Take the sup norm and the square $\{|x|\le\rho,\,|y|\le\rho\}$. The first component of the map only copies $y_n$, so $|x_{n+1}|\le\rho$ holds automatically. For the second component, write $G(y)=\alpha+\beta\cos(\omega\ln|y|)$ and $g(y)=y\,G(y)$, and let $M(\rho)=\max_{|y|\le\rho}|g(y)|$ be the farthest that the nonlinear term can throw a point of the square. Then $|y_{n+1}|\le b\rho+M(\rho)$, and the square is trapping whenever
\begin{equation}
\kappa(\rho)\equiv b+\frac{M(\rho)}{\rho}\le1 .
\label{eq:kappa}
\end{equation}
This condition is not tied to a single scale. Since $g(y)$ is magnified by $S$ when $y$ is, $M(S\rho)=S\,M(\rho)$, and so $\kappa$ is periodic in $\ln\rho$ with period $\ln S$. It holds in a window of radii in every period, so trapping squares exist at every scale, arbitrarily close to the origin included. Figure~\ref{fig:proof}a shows one of these squares. The origin is therefore Lyapunov stable, because orbits that start close to it stay close. It is not attracting, because the attractors $A_k$ also sit at every scale and orbits started near the origin settle on them instead.

A trapping ball is only useful if it is smaller than the attractor. A ball that contains $A_0$ tells us nothing about the basin of $A_0$, because orbits inside it can end on $A_0$. Here $A_0$ is any one attractor of the ladder, taken as the reference that fixes the labelling of the shells, and its size is $\ell_0=\max_{\mathbf{x}\in A_0}\|\mathbf{x}\|$, the largest distance from the origin reached on the attractor, in the same norm as the balls. The copy $A_k$ has size $S^{-k}\ell_0$. Whether a trapping ball smaller than the attractor exists depends on the nonlinearity and on the parameters, and the mechanism only needs one. We define $\rho^{*}_0$ as the largest trapping radius below $\ell_0$, the end of the last trapping window before the attractor, and $\rho^{*}_k=S^{-k}\rho^{*}_0$ for the rescaled copies, dashed lines in Fig.~\ref{fig:cascade}a. For the matryoshka map, $\ell_0=0.0773$ and $\rho^{*}_0=0.0341$, which is $0.44\,\ell_0$.

A further consequence of confinement is a bound on how far an orbit can reach. An orbit that starts inside a trapping ball never leaves it, so its limit set
lies inside the ball. We take the basin $B_k$ to be the set of initial conditions whose limit set is $A_k$, so an attractor that is not contained in the ball cannot be that limit. By covariance, $R(\rho^{*}_k)$ is trapping and $A_k$ is
not contained in it, so no point of the open ball $\|\mathbf{x}\|<\rho^{*}_k$ belongs to the basin $B_k$. Since this ball is open, it does not meet the
closure $\overline{B_k}$ either. Figure~\ref{fig:squares} sketches this for four consecutive shells. A boundary point at distance $r$ lies on the boundary of a basin $B_k$ only if it is a limit of points of $B_k$, which requires $r\ge\rho^{*}_k$. It therefore lies only on the boundaries of the basins with
\begin{equation}
k\;\ge\;\log_S\frac{\rho^{*}_0}{r},
\label{eq:reach}
\end{equation}
so the outermost basin it can reach is the one with the smallest integer $k$ satisfying Eq.~(\ref{eq:reach}), and all deeper ones (larger $k$, smaller attractors) may be present.

 Consider a boundary point at distance $r$ from the origin, and a shell $k$ whose trapping ball is larger than $r$, $\rho^{*}_k>r$, that is, $k<\log_S(\rho^{*}_0/r)$. Such shells exist for every $r$, because $\rho^{*}_k\to\infty$ as $k\to-\infty$. The point lies inside the ball $\|\mathbf{x}\|<\rho^{*}_k$, at a distance $\rho^{*}_k-r$ from its edge. Any disc around the point with a smaller radius lies entirely inside this ball, and the ball contains no point of $B_k$. So below this size no neighbourhood of the point contains points of $B_k$, and zooming in further does not change that. The point does not lie on the boundary of $B_k$. In a Wada boundary every point lies on the boundary of every basin, so no boundary point of the map is Wada.

The law also holds along orbits. If the orbit of $\mathbf{x}$ enters the open ball $\|\mathbf{x}\|<\rho^{*}_k$ at some step $n$, then by continuity of
$F^{n}$ a neighbourhood of $\mathbf{x}$ is mapped into this ball, where it stays trapped and cannot reach $A_k$. A point can therefore lie on the boundary of $B_k$ only if
its orbit never enters the ball,
\begin{equation}
\|F^{n}\mathbf{x}\|\;\ge\;\rho^{*}_k\qquad\text{for all } n\ge0 .
\label{eq:orbit}
\end{equation}
The case $n=0$ is Eq.~(\ref{eq:reach}). The other values of $n$ add conditions on the later iterates, so a point whose orbit passes close to the origin is excluded from the outer basins even if it starts far away.

The argument uses neither chaos, nor dissipation, nor the dimension, nor invertibility, only Eq.~(\ref{eq:cov}) and one trapping ball smaller than an attractor. Dissipation enters only through the existence of the attractors and not in the reach law. It is simple, and its consequences are not. Magnification lowers the reach of every point by one, because $\partial B_{k+1}=S^{-1}\partial B_k$, while the trapping balls forbid an unlimited reach outward. The boundaries inherit the symmetry toward the origin and not away from it.

The log-polar picture makes the law visible. Write a point as $\mathbf{x}=r\,\mathbf{n}(\theta)$, with $r=\|\mathbf{x}\|$ its distance from the origin and $\mathbf{n}(\theta)=(\cos\theta,\sin\theta)/\|(\cos\theta,\sin\theta)\|$ the unit vector of polar angle $\theta$, and use the logarithmic radius $u=\log_S r$ in place of $r$. A dilation $\mathbf{x}\to S\mathbf{x}$ then becomes the translation $u\to u+1$ at fixed $\theta$, so the pattern of basins repeats with period one, with $B_k$ carried into $B_{k-1}$ (Fig.~\ref{fig:cascade}a). In the figure $r$ is on a logarithmic axis, so $u\to u+1$ is a shift by $\log_{10}S\simeq0.87$ and the edge of $B_k$ sits at $r=\rho^{*}_k$. The basin $B_k$ has a vertical edge at $u=\log_S\rho^{*}_0-k$ and does not exist to its left. The magnifications in Fig.~\ref{fig:cascade}b,c show what this means at two boundary points. The point $p$ lies to the right of the edge of $B_{-2}$, so $B_{-2}$ is present there. The point $q$ lies to the left of that edge, inside the square of half-width $\rho^{*}_{-2}$ but outside the one of half-width $\rho^{*}_{-1}$, so its outermost basin is $B_{-1}$ and $B_{-2}$ never appears.

\subsection{The directed Wada cascade}

The results above give the structure that this paper is about. In a Wada boundary every point lies on the boundary of every basin, so all the basin boundaries
coincide. Here the boundaries are ordered instead. A point of reach $m(\mathbf{x})$ lies on the boundary of that basin and of no outer one, and, by the nesting in (i), of all the deeper ones. The
structure has the three properties below.
\begin{enumerate}
\item[(i)] Nesting. The boundary of every basin lies in the boundary of
every deeper one. A point that lies on the boundary of $B_k$ therefore lies on the boundary of every $B_{k'}$
with $k'\ge k$.
\item[(ii)] Finite reach. The reach of every boundary point is bounded
by its distance from the origin, $m(\mathbf{x})\ge\log_S(\rho^{*}_0/\|\mathbf{x}\|)$
(Eq.~\ref{eq:reach}), so no point lies on the boundary of all basins.
\item[(iii)] Strictness. The boundary of a basin differs from that of the next deeper one,
\begin{equation}
\partial B_k\neq\partial B_{k+1}=S^{-1}\partial B_k .
\label{eq:hierarchy}
\end{equation}
With (i), the deeper boundary has points that the outer one lacks, $\partial B_k\subsetneq\partial B_{k+1}$.
\end{enumerate}
Demagnifying a boundary by $S$ gives the boundary of the next deeper basin, which contains the original and magnifying it returns only a part. The structure is directed because the nesting runs one way, toward the origin the boundaries contain one another, but not away from it. The symmetry alone would allow both directions and the trapping balls forbid the outward one.

Property (ii) and the strictness in (iii) are the reach law, and they hold in every covariant system with a trapping ball. Strictness follows from (ii). If $\partial B_k=\partial B_{k+1}$ held for one $k$, scaling would make all the boundaries equal, and every boundary point would lie on the boundary of every basin, which (ii) forbids.

\begin{figure*}[t]
    \centering
    \includegraphics[width=\textwidth]{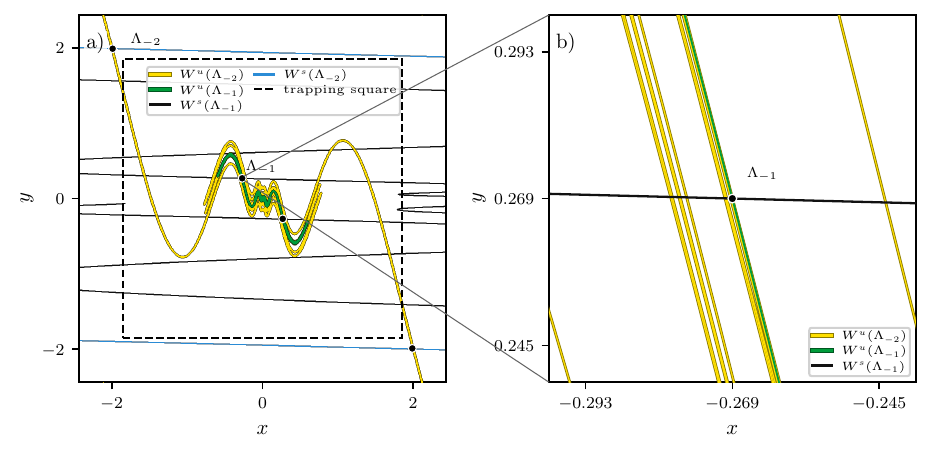}
    \caption{\textbf{Confinement and invariant manifolds.} Matryoshka map, in the view of shell $-1$. (a) The trapping square (dashed), of half-width $\simeq1.85$, holds the unstable manifold of $\Lambda_{-1}$ (green), while the stable manifold of $\Lambda_{-1}$ (black) crosses it. (b) Magnification around $\Lambda_{-1}$ (window of half-width $0.03$). The unstable manifold of $\Lambda_{-2}$ (yellow) crosses the stable manifold of $\Lambda_{-1}$ (black) transversally, and its strands accumulate on the unstable manifold of $\Lambda_{-1}$ (green). Unstable manifolds have the colour of their shell in Fig.~\ref{fig:cascade}.}
    \label{fig:proof}
\end{figure*}

The reach law says which basin boundaries a point cannot lie on. That it lies on the boundary of all the deeper ones, property (i), depends on the dynamics, and we obtain it from invariant manifolds. The matryoshka map is an invertible, odd map of the plane with a period-two saddle $\Lambda_k=S^{-k}\Lambda_0$ in each shell. We prove the nesting for the stable manifolds $W^s(\Lambda_k)$, which lie in the boundary of $B_k$, in four steps (Fig.~\ref{fig:proof}).

The first step is a crossing. A saddle has a stable manifold, the points that converge to it under iteration, and an unstable manifold, the points that converge to it under the inverse map. We need the unstable manifold of the saddle in one shell to cross the stable manifold of the saddle in the next, deeper shell, so that orbits leaving the first are carried to the second. This is what lets us use the lambda lemma in the next step. We prove the crossing with interval arithmetic at $300$ bits. Take a point $z_0$ of the saddle orbit $\Lambda_{-2}$ and its copy $z_1=S^{-1}z_0\in\Lambda_{-1}$. An arc of the local unstable manifold $W^u_{\rm loc}(z_0)$, of length $8\times10^{-10}$, is mapped by $F^{10}$ into a curve that crosses the local stable manifold $W^s_{\rm loc}(z_1)$ at exactly one point, transversally, at an angle of at least $73^\circ$ (Fig.~\ref{fig:proof}b). The method is the standard one for computer-assisted proofs of heteroclinic connections \cite{katokHasselblatt1995,wilczakZgliczynski2003,zgliczynskiGidea2004} (Appendix~\ref{app:proof}).

The second step is the nesting. The crossing means that $W^s(\Lambda_{k+1})$ cuts $W^u(\Lambda_k)$ transversally. The lambda lemma \cite{palisMorseSmaleDynamicalSystems1969}, applied to $F^{-2}$ near $\Lambda_k$, then says that the backward iterates of this curve accumulate on $W^s(\Lambda_k)$. Every point of $W^s(\Lambda_k)$ is therefore a limit of points of $W^s(\Lambda_{k+1})$, and by covariance this holds in every shell. Chaining the inclusions gives
\begin{equation}
W^s(\Lambda_k)\subset\overline{W^s(\Lambda_j)}\qquad(j>k).
\label{eq:wsnest}
\end{equation}

The third step is the horseshoe. The crossing of the first step joins two different saddles, $\Lambda_{-2}$ and $\Lambda_{-1}$, while the Smale--Birkhoff theorem needs a saddle whose unstable manifold returns to cross its own stable manifold. The dilation provides this. Since $F$ commutes with it, $F$ also acts on the points of the plane taken modulo dilation, where $\mathbf{x}$ and $S\mathbf{x}$ count as the same point. In the log-polar picture of Fig.~\ref{fig:cascade}a this means that $u$ and $u+1$ are identified, and we call the resulting map the reduced map $\bar F$. In $\bar F$ all the saddles $\Lambda_k$ are a single saddle $\bar\Lambda$, because they are copies of one another, and the crossing between neighbouring shells becomes a transversal homoclinic point of $\bar\Lambda$. By the Smale--Birkhoff theorem \cite{smale1965,palisTakens1993} this gives a horseshoe, an invariant Cantor set on which the dynamics is that of a symbol shift, and so chaos with positive topological entropy. The map is not smooth on the line $y=0$, but the theorem is local and the saddle and the certified orbit stay away from that line (Appendix~\ref{app:proof}).

The last step ties the manifolds to the basins. In interval arithmetic we build a trapping region $T$ around $A_0$ and show that the outward branch of the unstable manifold of $\Lambda_0$ enters it (Appendix~\ref{app:proof}). By the lambda lemma, every neighbourhood of a point of $W^s(\Lambda_0)$ then contains points that enter $T$ and so meet $B_0$, while the point itself is not in $B_0$, because it converges to the saddle. The stable manifold therefore lies in the boundary of $B_0$, and by covariance
\begin{equation}
\overline{W^s(\Lambda_k)}\subset\partial B_k .
\label{eq:wsB}
\end{equation}
Together with Eq.~(\ref{eq:wsnest}) it lies in $\partial B_j$ for every $j\ge k$. A point of $W^s(\Lambda_k)$ thus lies on the boundary of $B_k$ and of all deeper basins and, by Eq.~(\ref{eq:orbit}), of no outer one, its reach is exactly $k$. We have not proved the converse, that every point of $\partial B_k$ lies on $\overline{W^s(\Lambda_k)}$, which would give the nesting of property (i) for the whole boundary. It passes all the numerical tests we made, and we state it as an assumption.

As usually applied, the Wada diagnostics \cite{dazaTestingBasinsWada2015,dazaAscertainingWhenBasin2018} cannot tell the cascade from a Wada boundary. Near any boundary point there are infinitely many deeper basins, so every neighbourhood holds at least three of them and the tests report Wada at every distance. What the tests miss is that the outer basins are absent. They show it only if told where to look. Restricted to three consecutive shells, the merging test finds that the two deeper basins share their boundary, because merging them destroys $91.5\%$ of it, while merging any pair that includes the shallowest destroys $0.03\%$.

What does show the structure is the reach itself, as a function of distance. An experiment that prepares the amplitude $r$ and records the attractor it settles on finds a staircase (Fig.~\ref{fig:tests}), the largest attractor reached shrinks by $S$ each time $r$ shrinks by $S$. In a Wada system this envelope would be flat. The cells above the staircase are exactly empty, and the few empty cells below it are outcomes too rare to be sampled. A fraction of outcomes pooled over a window is a poor signature, because it is a log-periodic function of $r$ in every system with DSI, Wada or not, and it is dominated by the largest distance. The uncertainty is unbounded inward and bounded outward. From a distance $r$ the outcome can be any of the infinitely many deeper attractors, but never one larger than the reach allows.

 \section{A class of dilation-covariant maps}
\label{sec:class}

Any map that obeys Eq.~(\ref{eq:cov}) carries, with every attractor $A_0$, the ladder of copies $A_k=S^{-k}A_0$, and if it also has a trapping ball smaller than an attractor, it obeys the reach law of Sec.~\ref{sec:mech}. The question is how to build such maps. Two cases are excluded from the start. A linear map obeys Eq.~(\ref{eq:cov}) for every $S$, but it has no attractor other than the origin. More generally, a map that is homogeneous of degree one for every dilation, $F(\lambda\mathbf{x})=\lambda F(\mathbf{x})$ for all $\lambda>0$, has no preferred ratio and therefore no scale at which a bounded attractor could sit. The symmetry must therefore be discrete, so the nonlinearity has to be homogeneous of degree one for the ratio $S$ and its powers, and for no other ratio. The simplest way to get this is to multiply the linear term by a function that is periodic in the logarithm of the amplitude. Take
\begin{equation}
F(\mathbf{x})=L\mathbf{x}+W\,H(\mathbf{x})\,\mathbf{x},\qquad H(S\mathbf{x})=H(\mathbf{x}),
\label{eq:class}
\end{equation}
where $L$ and $W$ are constant matrices and $H(\mathbf{x})$ is a bounded matrix that does not change under a dilation, with $H(\mathbf{x})\mathbf{x}=0$ at the origin. Covariance is then immediate, $F(S\mathbf{x})=L\,S\mathbf{x}+W\,H(S\mathbf{x})\,S\mathbf{x}=S\,F(\mathbf{x})$, so Eq.~(\ref{eq:cov}) holds for any $L$, any $W$ and any such $H$. All the log-periodic modulation sits in $H$, and dilation invariance is the only thing asked of it. Three choices of $H$ cover our examples. The first is a diagonal $H$ in which each variable is multiplied by its own gain, $H=\mathrm{diag}\bigl[G(x_1),\dots,G(x_d)\bigr]$, where $x_1,\dots,x_d$ are the components of $\mathbf{x}$ and $G$ is a bounded function with $G(Ss)=G(s)$. A dilation of $x_j$ shifts $\ln|x_j|$ by one period and leaves its gain unchanged. The matryoshka map is of this type, with $W=\mathrm{diag}(0,1)$ so that only $y$ enters the nonlinearity, $G(s)=\alpha+\beta\cos(\omega\ln|s|)$ and $\ln S=2\pi/\omega$. The second is a gain that depends on the full radius, $H=G(\|\mathbf{x}\|)$ times the identity, which is the case of the radial log-periodic map. The third is a rotation by an angle proportional to $\ln\|\mathbf{x}\|$, which in complex notation is $z\,|z|^{i\omega}$ with $\omega\ln S=2\pi$, the case of the log-periodic rotation map. In logarithmic coordinates the dilation becomes a translation by $\ln S$, the counterpart of the additive symmetry of megastable systems \cite{sprott2017megastability,liInfinitescrollAttractorEnergy2016}. The same kind of log-periodic modulation has been used for the nonlinearity of flows in Lurie form
\cite{karimovExtremelyMultistableMatryoshka2025,burkinMatryoshkaMultistableSystem2026}, and Eq.~(\ref{eq:class}) is its counterpart for maps, with gains that may depend on several variables.

Covariance alone only guarantees a ladder of copies of whatever attractors exist. The cascade needs two further ingredients, an attractor and a trapping ball smaller than it. By covariance one trapping ball is enough, since it gives the whole ladder $S^{-k}\rho$ (Sec.~\ref{sec:mech}). What the log-periodic modulation provides is the first one. The farthest that the nonlinear term $W\,H(\mathbf{x})\,\mathbf{x}$ can throw a point of the ball $\|\mathbf{x}\|\le\rho$ is $M(\rho)$, its largest norm inside the ball. Because the gain is periodic in $\ln|x_j|$, $M(\rho)$ is set by the peak of the last period and stays fixed until the next peak, while $\rho$ keeps growing. Just above each peak, the ball is therefore larger than the image of its own contents, and if the linear part contracts, an orbit that enters it cannot leave. For the matryoshka map the condition is $\kappa(\rho)\le1$ of Eq.~(\ref{eq:kappa}), which holds in a window of radii in every period.

Trapping alone is also not enough. The orbits inside a trapping ball may simply collapse to the origin, and then there is nothing to confine. A ladder needs an attractor that stays away from the origin, which means that the drift of $\ln r$ along it must vanish, with stretching and contraction balancing each other. Without the trapping ball, in turn, the dynamics can carry an orbit outward and the direction of the cascade is lost. The cascade appears where both are present. Both are conditions on $L$, $W$ and the modulation, and not on the particular form of the nonlinearity, so we now show them in the matryoshka map and in four other systems that differ from it in many respects.

They are not automatic, and the examples below are chosen and not typical.

\section{Examples}
\label{sec:examples}

The reach law needs only Eq.~(\ref{eq:cov}) and a trapping ball, so we test it on systems that differ from the matryoshka map in as many other respects as possible, and not only on other parameters of the matryoshka map. The matryoshka map is the reference, we analyze it in full, and the nesting of the stable manifolds is proved for it. The radial log-periodic map lets the gain depend on the full radius, it is smooth and not invertible. The log-periodic rotation map enters the dilation as a rotation by an angle proportional to $\ln r$, and it has folds. The flow is a ring of three coupled variables. The network changes the dimension, with $N$ copies of the matryoshka map coupled through their average, so that $d=2N$ goes up to $128$. The norm is the sup norm, except for the log-periodic rotation map, where it is the Euclidean one. In every system we ask the same three questions. Does any initial condition end on an attractor larger than the reach law allows? How close to its threshold $\rho^{*}_k$ does each basin first appear, since it cannot appear before it? Does the largest reachable attractor step down by one shell for each factor $S$ in distance?

\begin{figure*}[t]
    \centering
    \includegraphics[width=0.8\textwidth]{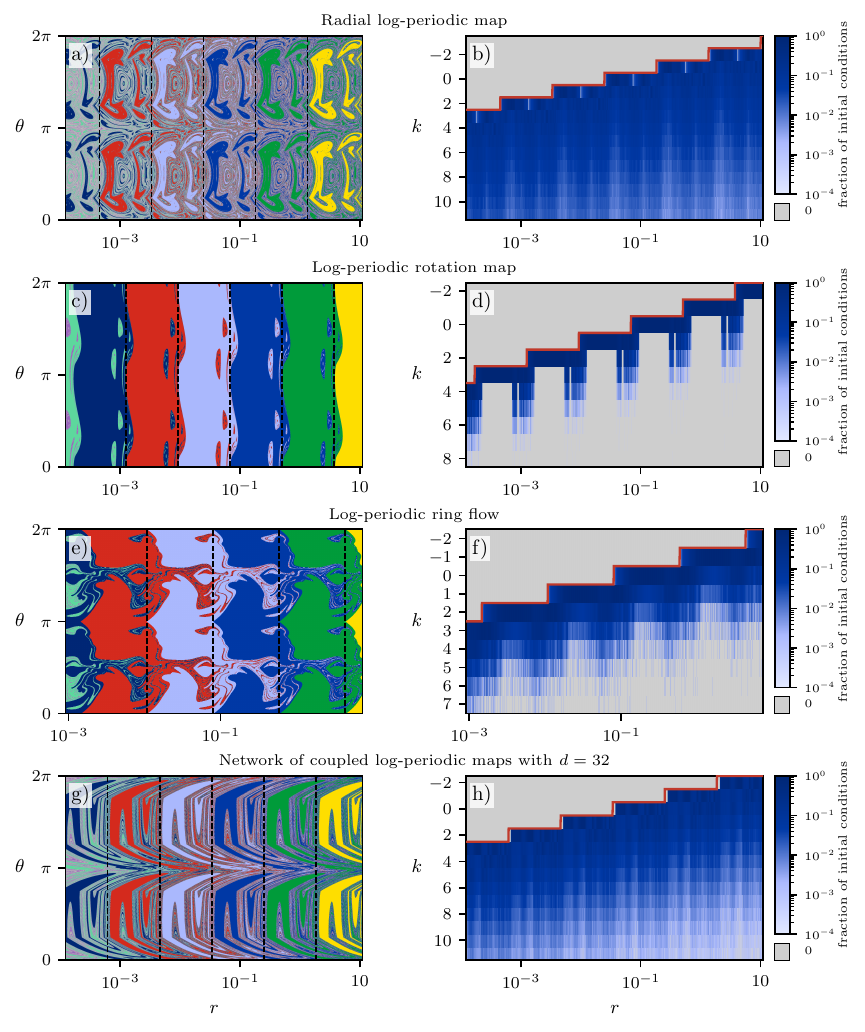}
    \caption{\textbf{The cascade beyond the matryoshka map.} The radial log-periodic map, the log-periodic rotation map, the log-periodic ring flow and the network of coupled log-periodic maps, one system per row. On the left, log-polar basin images coloured by the shell $k$, with dashed lines at the trapping thresholds. $r$ is the sup-norm distance, and the planes pass through the origin. For the flow the plane contains the diagonal $x_0=x_1=x_2$, and for the network ($d=32$) it is tilted by $0.1$ from the synchronous plane. On the right, for each radius, the fraction of the initial conditions that end on shell $k$ and the reach law (red staircase). The cells above the staircase are forbidden and never populated, and grey cells are those where no initial condition ended.}
    \label{fig:generality}
\end{figure*}

\subsection{Matryoshka map}
\label{sec:map1}

We start with the matryoshka map already presented, Eq.~(\ref{eq:map}). In the notation of Eq.~(\ref{eq:class}) it has $L=\begin{psmallmatrix}0&1\\-b&0\end{psmallmatrix}$, $W=\begin{psmallmatrix}0&0\\0&1\end{psmallmatrix}$ and gain $G(s)=\alpha+\beta\cos(\omega\ln|s|)$. This map is a homeomorphism of the plane, with inverse defined by $F^{-1}(x,y)=\bigl((g(x)-y)/b,\,x\bigr)$ and $g(x)=x\,G(x)$. Due to the use of a cosine function, it is an odd map, so attractors and basins come in symmetric pairs, and it is smooth except on the line $y=0$, where the derivative oscillates without a limit. Its Jacobian determinant is the constant $b$, so every region of the plane is contracted in area by $b$ at each step. The dissipation is thus fixed by $b$, through $L$, and the stretching by $\beta$, through the modulation.

Due to its construction the origin is non-hyperbolic and, for $\beta<|1+b-\alpha|$, is the only equilibrium. Confinement then makes every attractor hidden \cite{leonovLocalizationHiddenChuas2011,jafariElementaryQuadraticChaotic2013,wangNewTwoDimensionalMap2018}, and the open ball $\|\mathbf{x}\|<\rho^{*}_k$ around the origin contains no point of $B_k$, so no attractor $A_k$ can be reached by starting near the equilibrium.

The ladder is organized by the period-two cycles on the antidiagonal of the phase space. The point $(-\sigma,\sigma)$ returns to itself after two steps when the gain at amplitude $\sigma$ equals $-(1+b)$, that is, when $\cos(\omega\ln\sigma)=-(1+b+\alpha)/\beta$. Because the gain is periodic in $\ln\sigma$, there are solutions at every scale, a factor $S$ apart, but only if $\beta$ is large enough. Two families of cycles are born together in a saddle-node at
\begin{equation}
\beta_{c1}=|1+b+\alpha| ,
\label{eq:Bc1}
\end{equation}
and both are saddles. This is the mode locking, the log-amplitude returns to the same phase of the periodic gain after two steps instead of drifting through it. The family with strong contraction is the one that matters. Its stable manifold, with its copies in every shell, is part of the basin boundaries. One branch of its unstable manifold leads outward, to the chaotic attractor of its own shell, and the other leads inward and crosses the stable manifold of the saddle in the next shell. This heteroclinic connection is what makes the nesting of Sec.~\ref{sec:mech} run toward the origin. The other family bounds no basin.

Neither ingredient implies the other. Trapping squares exist in a window of radii in every period, with or without an attractor inside, and where there is no locking window the orbits inside them collapse to the origin. Where there is a ladder and no trapping square, outward boundary orbits appear and the direction is lost. The reach law holds in the matryoshka map with no violation, and each basin first appears at its threshold to within the resolution.

\subsection{Radial log-periodic map}
\label{sec:map2}

Two features of the matryoshka map could be changed, the line $y=0$, where it is not differentiable, and the constant Jacobian determinant. Both disappear if the gain depends on the full radius,
\begin{subequations}\label{eq:map2}
\begin{align}
x_{n+1}&=y_n,\\
y_{n+1}&=-b\,x_n+y_n\bigl[\alpha+\beta\cos(\omega\ln r_n)\bigr],\\
r_n&=\sqrt{x_n^2+y_n^2},
\end{align}
\end{subequations}
with $\alpha=-0.4$, $\beta=1.1$, $b=0.1$ and $\omega=\pi$, which we call the radial log-periodic map. It is covariant because a dilation shifts the argument of the cosine by $\omega\ln S=2\pi$, and it is smooth everywhere except at the origin. It is not invertible, and its Jacobian determinant varies from point to point and exceeds one near the boundary saddle, so it is not dissipative everywhere. The cascade does not depend on any of this. Trapping squares exist, each basin first appears at its threshold to within the resolution (Fig.~\ref{fig:generality}a,b), and the asymmetry between the boundaries is the same. Almost every boundary point of the outer basin $B_{-2}$ also lies on the boundary of $B_{-1}$, while only a small fraction of the boundary of $B_{-1}$ lies on the boundary of $B_{-2}$. The inward branch of the unstable manifold of the period-two saddle crosses the stable manifold of the next shell, and the generalized certificate, which needs $F$ to be injective only on the local boxes and not globally, proves two transversal heteroclinic orbits (Appendix~\ref{app:proof}). The nesting itself is numerical here.

\subsection{Log-periodic rotation map}
\label{sec:complex}

The log-periodic rotation map also has a simple formula, but its nonlinearity has no preferred axes. In the complex notation $z=x+iy$ it reads
\begin{equation}
  z_{n+1}=e^{i\varphi}\bigl(a\,z_n+\gamma\,z_n|z_n|^{i\omega}\bigr)+c\,\bar z_n ,
  \qquad \omega=\pi ,
  \label{eq:cmap}
\end{equation}
with $(a,\gamma,c,\varphi)=(0.75,0.7,0.3,1.2)$. The factor $|z|^{i\omega}=e^{i\omega\ln|z|}$ is a
rotation whose angle grows logarithmically with the radius. It is unchanged under $z\to Sz$,
$S=e^{2\pi/\omega}$, so $F(Sz)=S\,F(z)$ holds exactly. This is the rotation case of the class. The map is also odd and non-invertible, and the
Jacobian determinant takes both signs, so the map folds the plane.

The origin of the ladder is easiest to see at $c=0$. The angle then decouples and the modulus
obeys the closed one-dimensional map $r'=r\,|a+\gamma\,e^{i\omega\ln r}|$. Its gain oscillates
between $|a-\gamma|$ and $a+\gamma$, and a ladder of attracting circles exists whenever this interval
contains $1$. Within each period of $\ln r$ there is one attracting circle, at radii
$\simeq1.68\,S^{-k}$ for our parameters, and one repelling circle, at $\simeq0.60\,S^{-k}$. At $c=0$
every attracting circle has its own annular basin, bounded by repelling circles, so there is a
ladder but no cascade. The term $c\,\bar z$ couples angle and radius. It breaks the circles
into chaotic attractors ($\lambda\simeq0.22$) and the annuli into interleaved, folded basins.

The trapping condition is explicit. Since the angle of $z_n$ is free, the triangle inequality
is saturated, and a disc of radius $\rho$ is trapping if and only if
$\max_{r\le\rho}\,r\bigl(|a+\gamma\,r^{i\omega}|+c\bigr)\le\rho$. Together with covariance this
gives the reach law $k\ge\log_S(\rho_0^*/r)$ directly. Numerically (Fig.~\ref{fig:generality}c,d),
no point of any other basin lies inside the trapping ball of $A_k$ (these are hidden
attractors), points of other basins appear within about $10^{-2}$ in $\ln r$ outside it, and
the outermost shell reached at radius $r$ equals the bound for all but a fraction $\sim10^{-3}$
of radii, never falling below it. The outermost boundaries are nested in the directed way of the
other examples. Nearly all of $\partial B_{-2}$ lies on $\partial B_{-1}$, but only about a tenth
of $\partial B_{-1}$ lies on $\partial B_{-2}$.

\subsection{Log-periodic ring flow}
\label{sec:flow}

A flow needs the same two ingredients as a map. If $\dot{\mathbf{x}}=F(\mathbf{x})$ and $F(S\mathbf{x})=S\,F(\mathbf{x})$, a solution $\mathbf{x}(t)$ is taken into the solution $S\mathbf{x}(t)$ with the same time, because both sides of the equation are multiplied by $S$ and time is not rescaled. The time-$\tau$ map $\Psi_\tau$ then obeys $\Psi_\tau(S\mathbf{x})=S\,\Psi_\tau(\mathbf{x})$ for every $\tau$, and the ladder $A_k=S^{-k}A_0$ is present again, with the same time scales on every shell. In the plane such a ladder is made of limit cycles with annular basins, and each annulus shares a boundary only with its two neighbours, the intermixing needs a chaotic attractor, hence at least three variables. The simplest flow we found with all the ingredients, which we call the log-periodic ring flow, is a ring of three identical variables, each relaxing toward its predecessor through a log-periodic gain,
\begin{subequations}\label{eq:flow}
\begin{align}
\dot x_i&=-x_i+g(x_{i-1}),\qquad i=0,1,2\ (\mathrm{mod}\ 3),\\
g(x)&=x\,G(x),\qquad G(x)=\alpha+\beta\cos(\omega\ln|x|),
\end{align}
\end{subequations}
with $\omega=\pi$ and $(\alpha,\beta)=(1.2,\,2.05)$. The vector field has the form of Eq.~(\ref{eq:class}), with $L=-\mathbb{I}$, and $W=\begin{psmallmatrix}0&0&1\\1&0&0\\0&1&0\end{psmallmatrix}$, which feeds each variable with its predecessor. The gain $G$ is the same function in all three variables. The flow is odd, and it can be read as a three-stage ring oscillator whose amplifiers have a gain that oscillates with the logarithm of the amplitude. Since the diagonal of $W$ is zero, the nonlinear term does not contribute to the divergence, which is $-3$ at every point, so volumes contract at a rate fixed by the linear term alone, as the Jacobian determinant $b$ does in the matryoshka map.

Confinement is again exact and needs no integration. On a face $x_i=\rho$ of the cube $\max_i|x_i|\le\rho$ we have $\dot x_i\le-\rho+M(\rho)$, with $M(\rho)=\max_{|s|\le\rho}|g(s)|$, so the cube is trapping when $M(\rho)\le\rho$. By covariance $M(\rho)/\rho$ is periodic in $\ln\rho$, and it is below one in a window $\rho\in[0.48,0.59]\,S^{-k}$ in every period, about a tenth of it. The reach law is then the same statement as for the maps, with cubes in place of squares.

The skeleton of the ladder is again a set of equilibria, one family per scale, and it is simpler than the period-two cycles of the matryoshka map because it sits on the diagonal $x_0=x_1=x_2$, which is invariant by the cyclic symmetry. On it the flow is one-dimensional, $\dot x=-x+g(x)$, and its equilibria, $g(x)=x$, come in two families per period, at $0.23\,S^{-k}$ and $0.59\,S^{-k}$. The first is stable along the diagonal and a spiral source in the transverse plane, with eigenvalues $-6.4$ and $1.7\pm4.7i$, and the chaotic attractor of the shell winds around it. The second is unstable along the diagonal and attracting in the transverse plane, with eigenvalues $6.4$ and $-4.7\pm6.4i$, a saddle-focus whose two-dimensional stable manifold, with its copies in every shell, is where the basin boundaries pass, and its one-dimensional unstable manifold is the diagonal, which runs from one shell to the next. This plays the role of the boundary saddle of the matryoshka map. Along the diagonal itself the basin boundaries sit exactly at these equilibria, and the threshold radii of the reach law coincide with them, so the trapping window ends where the saddle-focus is.

The tests are those of the other systems. A set of $6\times10^{4}$ initial conditions in three dimensions was used and none ended on an attractor larger than the reach law allows. Figure~\ref{fig:generality}e,f show the basins on a plane through the origin that contains the diagonal. The bands of the shells are crossed by fingers of the neighbouring basins, as in the maps, the bound is never violated, each basin appears at its threshold to within the resolution, and the largest reachable shell steps down by one for each factor $S$ in distance (Fig.~\ref{fig:generality}f). The asymmetry between the boundaries is the same as in the maps, most of the boundary of a basin also lies on the boundary of the basin of the next, smaller attractor, and much less the other way. The classification of the final state changes by about one percent when the time step is divided by four, and these are points near the basin boundaries. The ladder and the confinement are exact, the chaos is numerical.

\subsection{Network of coupled matryoshka maps}
\label{sec:network}

The last example changes the dimension. We take $N$ identical copies of the matryoshka map, with states $(x_i,y_i)$, and let them interact through a global coupling of strength $\varepsilon$. At each step every copy is first updated by the matryoshka map, and the result is then mixed with the average over all copies,
\begin{subequations}\label{eq:net}
\begin{align}
x_{i,n+1}&=(1-\varepsilon)\,y_{i,n}+\varepsilon\,\langle y\rangle_n,\\
y_{i,n+1}&=(1-\varepsilon)\,\bigl[-b\,x_{i,n}+g(y_{i,n})\bigr]+\varepsilon\,\bigl\langle -b\,x+g(y)\bigr\rangle_n,
\end{align}
\end{subequations}
with $g(y)=y\,G(y)$ as in the matryoshka map, and the dimension of the phase space is $d=2N$. For $\varepsilon=0$ the copies are independent, and for $\varepsilon=1$ they all receive the same input. The coupling is linear, so it commutes with the dilation and the network is covariant, $F_{\rm net}(S\mathbf{x})=S\,F_{\rm net}(\mathbf{x})$. We use the parameters of the matryoshka map and $\varepsilon=0.8$, with $d$ up to $128$.

Confinement is inherited from the matryoshka map, and this is why the example is useful. The reach law should not care about the dimension, and here it can be tested in dimensions where nothing else is simple. An average with non-negative weights that add to one cannot increase the sup norm, so a cube of half-width $\rho$ is trapping for the network exactly when the square of the same size is trapping for the matryoshka map, that is, when $\kappa(\rho)\le1$ in Eq.~(\ref{eq:kappa}). The thresholds $\rho^{*}_k$ are those of the matryoshka map for any $N$, any $\varepsilon$ and any non-negative coupling matrix with rows that add to one, rings and random graphs included. What can change with $d$ is the intermixing of the basins, and that is what we test.

The state with all copies equal is invariant, and on it the network is the matryoshka map, so the ladder of attractors is in the network. These attractors are stable against differences between the copies for $\varepsilon>0.29$, and the boundary saddle of the matryoshka map, which carries the nesting, keeps its single unstable direction for $\varepsilon>0.74$, and for $\varepsilon=0.8$ its two branches lead to the shells $k$ and $k+1$ as in the matryoshka map, so the skeleton of the cascade is inherited. We did not repeat the interval-arithmetic certificate for the network, and the directed structure is numerical.

For the reach law we tested $14$ networks that differ in topology, dimension and coupling, namely global coupling with $d=4$ to $128$ and $\varepsilon$ between $0.3$ and $1$, rings with $d=8$ to $32$, and a random $4$-regular graph with $d=32$. Each was started from isotropic initial conditions, from initial conditions with random anisotropy and from a single excited copy, about $3\times10^{5}$ states per network and $4\times10^{6}$ in total, and none ended on an attractor larger than the reach law allows.

Figures~\ref{fig:generality}g,h show the basins for $d=32$ on a plane through the origin tilted by $0.1$ from the synchronous plane. The bands of the shells, the fingers of the neighbouring basins that cross them and the staircase of the largest attractor are those of the matryoshka map. From a given distance, the largest attractor that is reached sits exactly at the bound in $96\%$ of the radii. On a random plane through the origin the staircase is the same but shifted by one shell toward smaller attractors, because a generic plane stays away from the directions near the synchronous plane, which are the ones that reach the bound.

What changes with the dimension is how often the bound is reached. For initial conditions drawn isotropically at a given distance, the fraction that end on the largest attractor the bound allows falls from $0.39$ at $d=2$ to $0.008$ at $d=32$ and is below $10^{-3}$ for $d\ge64$, and a typical initial condition ends two or three shells beyond it. The bound stays exact and stays a limit, but it is saturated only by the initial conditions close to the synchronous states, a set that shrinks with $d$. The directed asymmetry goes the other way. On random planes, the fraction of the boundary of $B_0$ that lies on the boundary of $B_1$ stays near $0.97$ up to $d=32$ and is $0.92$ at $d=64$, while the fraction of the boundary of $B_1$ that lies on the boundary of $B_0$ falls from $0.48$ to $0.21$, so the cascade becomes more one-way as the dimension grows. These fractions are measured on two-dimensional slices.

\section{Conclusions}
\label{sec:disc}

We started from a simple question. If the equations of motion have discrete scale invariance, a magnification carries each attractor onto another, and the system holds an infinite ladder of attractors whose basins are exact copies of one another. What does this symmetry do to the basin boundaries? A discrete translation passes the symmetry to the boundaries in full. A magnification and its inverse are treated alike by the dynamics and the ladder has no preferred shell, so the natural expectation was a Wada boundary, in which every boundary point lies on the boundary of every basin.

The expectation fails. Dilation covariance together with a single trapping region fixes an exact bound on which basin boundaries a point can lie on, the reach law. It uses neither chaos, nor dissipation, nor invertibility, nor a particular dimension, and dissipation is needed only for the attractors to exist. No point lies on the boundary of all of the basins, so no boundary point is Wada, and the basins on whose boundary a point does lie are the deeper ones, as proved for the stable manifolds and assumed for the rest of the boundary. The boundaries are nested in one direction only, toward the origin, a directed Wada cascade. The symmetry is split in two because the origin that collects the copies of the attractors also anchors the trapping regions, which exist at every scale. A magnification lowers the reach of every point by one shell, while the trapping regions forbid an unlimited reach outward. In the matryoshka map we proved that the stable manifolds of the boundary saddles are nested in this way, through a heteroclinic connection between neighbouring shells. The mechanism is not tied to this map. A log-periodic modulation of a homogeneous nonlinearity supplies the ladder of attractors, through mode locking, and the ladder of trapping regions, through the same periodicity, and we found the cascade in two other maps, in a flow and in a high-dimensional network.

For predictability the consequence is an asymmetry in the final-state uncertainty. From a distance $r$ to the origin the outcome can be any of infinitely many attractors, but only those below a limit set by $r$, so the uncertainty is unbounded inward and bounded outward. As usually applied, the standard Wada diagnostics cannot see this, because they ask only how many basins meet near a point and there are always at least three. They have to be complemented by the question that the symmetry suggests, on which basin boundaries a point lies as a function of its distance. This is also what an experiment measures, by preparing an amplitude and recording where the system settles. We expect the same structure wherever discrete scale invariance acts on a dissipative dynamics, and not only on a spectrum or a scaling law.

Several questions remain open. The nesting of the stable manifolds is proved for the matryoshka map, up to the numerical check that the trapping region around an attractor holds no other attractor, but we did not prove that these manifolds make up the whole boundary, and we state it as an assumption supported by the numerics. For the radial log-periodic map only the crossing of the manifolds is certified, and for the log-periodic rotation map, the flow and the network the directed structure is numerical. Extending the certificates to these cases is the natural next step. In high dimension the bound is saturated by fewer and fewer initial conditions while the asymmetry between the boundaries grows, and understanding this trend is another. The last is a physical realization of the ladder, for instance in an electronic circuit whose gain oscillates with the logarithm of the amplitude, where the reach law could be tested directly.

\acknowledgments

P. Haerter and I. L. Caldas acknowledge the São Paulo Research Foundation (FAPESP), Brasil, under Process Numbers 2025/28656-4, 2023/16146-6, and 2024/05700-5, respectively. R. L. Viana was supported by CNPq (Brazil) under Grants No. 403120/2021-7 and No. 301019/2019-3. This research was also supported by INCT-NeuroComp (CNPq Grant No. 408389/2024-9). M. A. F. Sanjuán acknowledges financial support from the Spanish State Research Agency (AEI) and the European Regional Development Fund (ERDF, EU) under Project No.~PID2023-148160NB-I00 (MCIN/AEI/10.13039/501100011033).

\appendix

\section{Computer-assisted proofs and numerical checks}
\label{app:proof}

A computer-assisted proof replaces each real number in a calculation by an interval that is guaranteed to contain it, with the rounding errors included in the bounds. Every inequality that the program reports then holds for all the points inside the intervals, and not only for one floating-point value. We use this to check a finite list of inequalities which, together with standard theorems that we cite and do not reprove, imply the statements below. The arithmetic is done with Arb \cite{johansson2017arb} through python-flint \cite{pythonflint} at $300$ bits. The code was not formally verified, so a certificate is as reliable as the program that produced it. Whatever is not certified is called numerical.

\subsection{The crossing between neighbouring shells}
\label{app:cross}

The result is that, for the matryoshka map with $\alpha=-1/2$, $\beta=27/25$, $b=1/10$ and $\omega=\pi$, there is a point $\xi$ in the inward branch of the local unstable manifold $W^u_{\rm loc}(z_0)$ of a saddle $z_0$ such that $F^{10}(\xi)$ lies in the local stable manifold $W^s_{\rm loc}(z_1)$ of its copy $z_1=S^{-1}z_0$ in the next shell. The point $\xi$ is unique in the arc examined, the manifolds are transversal there, and the orbit of $\xi$ keeps $|y|\ge0.269$, away from the line $y=0$ where the map is not smooth.

\begin{table}[b]
\caption{Certificate of the crossing in the matryoshka map. The coordinates $(u',s')$ are those at $z_1$ and $\delta_1=\delta_0/S$.}
\label{tab:cert}
\begin{ruledtabular}
\small
\begin{tabular}{ll}
Quantity & Value \\
\hline
Saddle position $\sigma$ & $1.98872629507339$ \\
Multipliers of $D\Phi(z_0)$ & $15.1747$, $6.59\times10^{-4}$ \\
Box and cone slope & $\delta_0=10^{-3}$, $\mu=5\times10^{-3}$ \\
Iterations & $10$ \\
Ends of the image in $u'$ & $-2.7063\times10^{-4}$, $+2.7071\times10^{-4}$ \\
Box half-width $\delta_1$ & $1.353\times10^{-4}$ \\
Image in $s'$ & $|s'|\le2.364\times10^{-5}$ \\
Slope of the image & $\le4.19\times10^{-4}$ \\
Smallest $|y|$ along the orbit & $\ge0.2691$ \\
\end{tabular}
\end{ruledtabular}
\end{table}

The saddle $z_0=(-\sigma,\sigma)$ is known in closed form, with $\cos(\omega\ln\sigma)=-(1+b+\alpha)/\beta$. Let $\Phi=F^{2}$ and let $(u,s)$ be coordinates centred at $z_0$ along the unstable and stable directions of $D\Phi(z_0)$. In the box $|u|,|s|\le\delta_0=10^{-3}$ we verify with intervals that $\Phi$ maps the cone $|s|\le\mu|u|$ into itself, that $\Phi^{-1}$ maps the cone $|u|\le\mu|s|$ into itself, and that $\Phi$ expands along $u$ and contracts along $s$, with $\mu=5\times10^{-3}$. The graph-transform theorem \cite{katokHasselblatt1995} then gives $W^u_{\rm loc}$ and $W^s_{\rm loc}$ as $C^1$ graphs of slope at most $\mu$. By covariance the same holds at $z_1$, in the box of half-width $\delta_1=\delta_0/S$.

An arc of $W^u_{\rm loc}(z_0)$ of length $8\times10^{-10}$ is then mapped ten times by $F$. At each step its image is enclosed in a thin parallelogram carried by the derivative of $F$, plus a box that bounds the remainder, and the enclosure must exclude $y=0$. After ten steps the arc is stretched by a factor $6.5\times10^{5}$. Its two ends lie on opposite sides of the box around $z_1$, and the whole image stays inside the box along the stable direction (Table~\ref{tab:cert}). The image is a graph over the unstable coordinate of slope at most $4.19\times10^{-4}$, and $W^s_{\rm loc}(z_1)$ is a graph over the stable coordinate of slope at most $\mu$. Their product is below one, so the composition of the two graphs is a contraction and the curves meet at exactly one point. The eigendirections make an angle of $74.1^\circ$ and the curves tilt away from them by less than $0.4^\circ$ in total, so the crossing is transversal at an angle of at least $73^\circ$.

Two checks that do not enter the proof agree with this. Random points of the initial arc, mapped ten times in ball arithmetic, all land inside the final enclosure, and the slope of the image of the centre line, computed by high-precision finite differences, lies inside the certified bound.

\subsection{The stable manifold lies in the basin boundary}
\label{app:basin}

This section proves Eq.~(\ref{eq:wsB}) for $\Lambda_0$ and so, by covariance, for every shell. It needs a trapping region $T$ around $A_0$ that the outward branch of the unstable manifold of $\Lambda_0$ enters.

The region $T$ is a union of $401$ slabs, found in double precision and enlarged until it was closed under the map. It is proved trapping by verifying in Arb that the image of each of $11\,435$ pieces lies in its interior. A square of half-width $0.15$ is trapping ($\kappa\le0.67$), contains $T$ and does not meet $S\,T$, which keeps the larger copies of $A_0$ out of it. Assuming that $A_0$ is the only attractor inside $T$, which we checked numerically and did not certify, $T$ lies in $B_0$. The saddle $\Lambda_0$ has norm $0.0364$, below the smallest norm in $T$, $0.0403$, so it lies outside $T$.

Let $x$ be a point of $W^s(\Lambda_0)$ whose orbit avoids $y=0$, so that $F$ is smooth along it. By the lambda lemma applied to $F^{2}$ at the saddle, the iterates of a small disc around $x$ contain a curve $C^1$-close to the outward branch. This curve enters $T$, so the disc meets $B_0$ and $x$ lies in $\overline{B_0}$. Such points are dense in $W^s(\Lambda_0)$. No point of $W^s(\Lambda_0)$ belongs to $B_0$, because an orbit in $B_0$ enters $T$ and stays there, while an orbit in $W^s(\Lambda_0)$ converges to $\Lambda_0$, which is not in $\overline{T}$. Since $B_0$ is open, $\overline{W^s(\Lambda_0)}\subset\overline{B_0}\setminus B_0=\partial B_0$, which with covariance is Eq.~(\ref{eq:wsB}).

\bibliography{Refs}

@article{aguirreFractalStructuresNonlinear2009,
  title     = {Fractal Structures in Nonlinear Dynamics},
  author    = {Aguirre, Jacobo and Viana, Ricardo L. and Sanju{\'a}n, Miguel A. F.},
  year      = 2009,
  month     = mar,
  journal   = {Rev. Mod. Phys.},
  volume    = {81},
  number    = {1},
  pages     = {333--386},
  publisher = {American Physical Society},
  doi       = {10.1103/RevModPhys.81.333},
  urldate   = {2023-09-17}
}

@article{HowTodetecWadabasins,
  title    = {How to detect Wada basins},
  journal  = {Discrete and Continuous Dynamical Systems - B},
  volume   = {26},
  number   = {1},
  pages    = {717-739},
  year     = {2021},
  issn     = {1531-3492},
  doi      = {10.3934/dcdsb.2020330},
  url      = {https://www.aimsciences.org/article/id/be5beb54-93b6-4c61-80c1-d815ece8ef92},
  author   = {Alexandre Wagemakers and Alvar Daza and Miguel A. F. Sanjuán}
}

@article{karimov2024matryoshka,
  title     = {Matryoshka multistability: Coexistence of an infinite number of exactly self-similar nested attractors in a fractal phase space},
  author    = {Karimov, Artur and Babkin, Ivan and Rybin, Vyacheslav and Butusov, Denis},
  journal   = {Chaos Solitons Fractals},
  volume    = {187},
  pages     = {115412},
  year      = {2024},
  doi       = {10.1016/j.chaos.2024.115412},
  publisher = {Elsevier}
}

@article{sprott2017megastability,
  title     = {Megastability: Coexistence of a countable infinity of nested attractors in a periodically-forced oscillator with spatially-periodic damping},
  author    = {Sprott, Julien C. and Jafari, Sajad and Khalaf, Abdul Jalil M. and Kapitaniak, Tomasz},
  journal   = {Eur. Phys. J. Spec. Top.},
  volume    = {226},
  number    = {9},
  pages     = {1979--1985},
  year      = {2017},
  doi       = {10.1140/epjst/e2017-70037-1},
  publisher = {Springer}
}

@article{dazaAscertainingWhenBasin2018,
  title     = {Ascertaining when a basin is {{Wada}}: the merging method},
  author    = {Daza, Alvar and Wagemakers, Alexandre and Sanju{\'a}n, Miguel A. F.},
  journal   = {Sci. Rep.},
  volume    = {8},
  number    = {1},
  pages     = {9954},
  year      = {2018},
  publisher = {Nature Publishing Group UK London}
}

@article{braatenUniversalityFewbodySystems2006,
  title     = {Universality in Few-Body Systems with Large Scattering Length},
  author    = {Braaten, Eric and Hammer, H.-W.},
  year      = 2006,
  month     = jun,
  journal   = {Phys. Rep.},
  volume    = {428},
  number    = {5-6},
  pages     = {259--390},
  issn      = {03701573},
  doi       = {10.1016/j.physrep.2006.03.001},
  urldate   = {2026-06-23},
  copyright = {https://www.elsevier.com/tdm/userlicense/1.0/},
  langid    = {english}
}

@article{dazaMultistabilityUnpredictability2024,
  title   = {Multistability and Unpredictability},
  author  = {Daza, {\'A}lvar and Wagemakers, Alexandre and Sanju{\'a}n, Miguel A. F.},
  year    = 2024,
  journal = {Phys. Today},
  volume  = {77},
  pages   = {44--50}
}

@article{jafariElementaryQuadraticChaotic2013,
  title    = {Elementary Quadratic Chaotic Flows with No Equilibria},
  author   = {Jafari, Sajad and Sprott, J. C. and Hashemi Golpayegani, S. Mohammad Reza},
  year     = 2013,
  month    = mar,
  journal  = {Phys. Lett. A},
  volume   = {377},
  number   = {9},
  pages    = {699--702},
  issn     = {0375-9601},
  doi      = {10.1016/j.physleta.2013.01.009},
  urldate  = {2026-04-27}
}

@article{johansenNewEvidenceEarthquake2000,
  title     = {New Evidence of Earthquake Precursory Phenomena in the 17 {{January}} 1995 {{Kobe}} Earthquake, {{Japan}}},
  author    = {Johansen, A. and Saleur, H. and Sornette, D.},
  year      = 2000,
  month     = may,
  journal   = {Eur. Phys. J. B},
  volume    = {15},
  number    = {3},
  pages     = {551--555},
  issn      = {1434-6028},
  doi       = {10.1007/s100510051159},
  urldate   = {2026-06-23},
  copyright = {http://www.springer.com/tdm},
  langid    = {english}
}

@article{karimovExtremelyMultistableMatryoshka2025,
  title     = {Extremely Multistable Matryoshka Systems with Hidden Attractors},
  author    = {Karimov, Artur and Rybin, Vyacheslav and Butusov, Denis and Kuznetsova, Oksana and Burkin, Igor},
  year      = 2025,
  month     = dec,
  journal   = {Chaos},
  volume    = {35},
  number    = {12},
  pages     = {123129},
  publisher = {AIP Publishing},
  issn      = {1054-1500},
  doi       = {10.1063/5.0289328},
  urldate   = {2026-04-27},
  langid    = {english},
  pmid      = {41396026}
}

@article{kennedyBasinsWada1991,
  title    = {Basins of {{Wada}}},
  author   = {Kennedy, Judy and Yorke, James A.},
  year     = 1991,
  month    = aug,
  journal  = {Physica D},
  volume   = {51},
  number   = {1},
  pages    = {213--225},
  issn     = {0167-2789},
  doi      = {10.1016/0167-2789(91)90234-Z},
  urldate  = {2023-09-20}
}

@article{kraemerEvidenceEfimovQuantum2006,
  title     = {Evidence for {{Efimov}} Quantum States in an Ultracold Gas of Caesium Atoms},
  author    = {Kraemer, T. and Mark, M. and Waldburger, P. and Danzl, J. G. and Chin, C. and Engeser, B. and Lange, A. D. and Pilch, K. and Jaakkola, A. and N{\"a}gerl, H.-C. and Grimm, R.},
  year      = 2006,
  month     = mar,
  journal   = {Nature},
  volume    = {440},
  number    = {7082},
  pages     = {315--318},
  publisher = {Nature Publishing Group},
  issn      = {1476-4687},
  doi       = {10.1038/nature04626},
  urldate   = {2026-06-23},
  copyright = {2006 Springer Nature Limited},
  langid    = {english}
}

@article{leonovLocalizationHiddenChuas2011,
  title    = {Localization of Hidden {{Chua's}} Attractors},
  author   = {Leonov, G. A. and Kuznetsov, N. V. and Vagaitsev, V. I.},
  year     = 2011,
  month    = jun,
  journal  = {Phys. Lett. A},
  volume   = {375},
  number   = {23},
  pages    = {2230--2233},
  issn     = {0375-9601},
  doi      = {10.1016/j.physleta.2011.04.037},
  urldate  = {2026-04-27}
}

@article{nusseWadaBasinBoundaries1996,
  title    = {Wada Basin Boundaries and Basin Cells},
  author   = {Nusse, Helena E. and Yorke, James A.},
  year     = 1996,
  month    = feb,
  journal  = {Physica D},
  volume   = {90},
  number   = {3},
  pages    = {242--261},
  issn     = {0167-2789},
  doi      = {10.1016/0167-2789(95)00249-9},
  urldate  = {2023-09-20}
}

@article{palisMorseSmaleDynamicalSystems1969,
  title     = {On {{Morse-Smale}} Dynamical Systems},
  author    = {Palis, J.},
  year      = 1969,
  month     = sep,
  journal   = {Topology},
  volume    = {8},
  number    = {4},
  pages     = {385--404},
  issn      = {00409383},
  doi       = {10.1016/0040-9383(69)90024-X},
  urldate   = {2026-06-23},
  copyright = {https://www.elsevier.com/tdm/userlicense/1.0/},
  langid    = {english}
}

@article{sornetteDiscretescaleInvarianceComplex1998,
  title     = {Discrete-Scale Invariance and Complex Dimensions},
  author    = {Sornette, Didier},
  year      = 1998,
  month     = apr,
  journal   = {Phys. Rep.},
  volume    = {297},
  number    = {5},
  pages     = {239--270},
  issn      = {03701573},
  doi       = {10.1016/S0370-1573(97)00076-8},
  urldate   = {2026-06-23},
  copyright = {https://www.elsevier.com/tdm/userlicense/1.0/},
  langid    = {english}
}

@article{wangDiscoveryLogperiodicOscillations2018,
  title    = {Discovery of Log-Periodic Oscillations in Ultraquantum Topological Materials},
  author   = {Wang, Huichao and Liu, Haiwen and Li, Yanan and Liu, Yongjie and Wang, Junfeng and Liu, Jun and Dai, Ji-Yan and Wang, Yong and Li, Liang and Yan, Jiaqiang and Mandrus, David and Xie, X. C. and Wang, Jian},
  year     = 2018,
  month    = nov,
  journal  = {Sci. Adv.},
  volume   = {4},
  number   = {11},
  pages    = {eaau5096},
  issn     = {2375-2548},
  doi      = {10.1126/sciadv.aau5096},
  urldate  = {2026-06-23},
  langid   = {english}
}

@article{wangInfiniteNumberWada2023,
  title   = {Infinite Number of {{Wada}} Basins in a Megastable Nonlinear Oscillator},
  author  = {Wang, Jingwei and Zhang, Yongxiang},
  year    = 2023,
  month   = jun,
  journal = {Nonlinear Dyn.},
  volume  = {111},
  number  = {11},
  pages   = {10601--10615},
  issn    = {0924-090X, 1573-269X},
  doi     = {10.1007/s11071-023-08365-2},
  urldate = {2026-07-01},
  langid  = {english}
}

@article{wangLogperiodicQuantumMagnetooscillations2019,
  title     = {Log-Periodic Quantum Magneto-Oscillations and Discrete-Scale Invariance in Topological Material {{HfTe5}}},
  author    = {Wang, Huichao and Liu, Yanzhao and Liu, Yongjie and Xi, Chuanying and Wang, Junfeng and Liu, Jun and Wang, Yong and Li, Liang and Lau, Shu Ping and Tian, Mingliang and Yan, Jiaqiang and Mandrus, David and Dai, Ji-Yan and Liu, Haiwen and Xie, Xincheng and Wang, Jian},
  year      = 2019,
  month     = oct,
  journal   = {Natl. Sci. Rev.},
  volume    = {6},
  number    = {5},
  pages     = {914--920},
  issn      = {2095-5138, 2053-714X},
  doi       = {10.1093/nsr/nwz110},
  urldate   = {2026-06-23},
  copyright = {http://creativecommons.org/licenses/by/4.0/},
  langid    = {english}
}

@article{wangNewTwoDimensionalMap2018,
  title     = {A new two-dimensional map with hidden attractors},
  author    = {Wang, Chuanfu and Ding, Qun},
  year      = 2018,
  month     = may,
  journal   = {Entropy},
  volume    = {20},
  number    = {5},
  pages     = {322},
  publisher = {Multidisciplinary Digital Publishing Institute},
  issn      = {1099-4300},
  doi       = {10.3390/e20050322},
  urldate   = {2026-04-27},
  copyright = {http://creativecommons.org/licenses/by/3.0/},
  langid    = {english}
}

@article{zhouNEWEVIDENCEDISCRETE2003,
  title    = {New evidence of discrete scale invariance in the energy dissipation of three-dimensional turbulence: Correlation approach and direct spectral detection},
  author   = {Zhou, Wei-Xing and Sornette, Didier and Pisarenko, Vladilen},
  year     = 2003,
  month    = may,
  journal  = {Int. J. Mod. Phys. C},
  volume   = {14},
  number   = {04},
  pages    = {459--470},
  issn     = {0129-1831, 1793-6586},
  doi      = {10.1142/S0129183103004632},
  urldate  = {2026-07-01},
  langid   = {english}
}

@article{liInfinitescrollAttractorEnergy2016,
  title   = {The Infinite-Scroll Attractor and Energy Transition in Chaotic Circuit},
  author  = {Li, Fan and Yao, Chenggui},
  journal = {Nonlinear Dyn.},
  volume  = {84},
  pages   = {2305--2315},
  year    = {2016},
  doi     = {10.1007/s11071-016-2646-z}
}

@article{zaccantiObservationEfimovSpectrum2009,
  title   = {Observation of an {E}fimov spectrum in an atomic system},
  author  = {Zaccanti, M. and Deissler, B. and D'Errico, C. and Fattori, M. and Jona-Lasinio, M. and M{\"u}ller, S. and Roati, G. and Inguscio, M. and Modugno, G.},
  journal = {Nat. Phys.},
  volume  = {5},
  pages   = {586--591},
  year    = {2009}
}

@article{dazaTestingBasinsWada2015,
  author  = {Daza, Alvar and Wagemakers, Alexandre and Sanju{\'a}n, Miguel A. F. and Yorke, James A.},
  title   = {Testing for Basins of {{Wada}}},
  journal = {Sci. Rep.},
  volume  = {5},
  pages   = {16579},
  year    = {2015},
  doi     = {10.1038/srep16579}
}

@article{zhangWadaBifurcationsPartially2013,
  author  = {Zhang, Yongxiang and Luo, Guanwei},
  title   = {Wada bifurcations and partially {W}ada basin boundaries in a two-dimensional cubic map},
  journal = {Phys. Lett. A},
  volume  = {377},
  pages   = {1274--1281},
  year    = {2013},
  doi     = {10.1016/j.physleta.2013.03.027}
}

@article{wangFractalInducedMultistabilityChuas2025,
  author  = {Wang, Ning and Lv, Yifeng and Bao, Han and Zhao, Shuhao and Xu, Quan},
  title   = {Fractal-induced multistability in {C}hua's circuit},
  journal = {Chaos Solitons Fractals},
  volume  = {200},
  pages   = {116939},
  year    = {2025},
  doi     = {10.1016/j.chaos.2025.116939}
}

@misc{xieSymmetryGroupsBasins2025,
  author       = {Xie, Xiao},
  title        = {Symmetry Groups of Basins of Attraction in Equivariant Dynamical Systems},
  howpublished = {arXiv:2509.11403},
  year         = {2025}
}

@incollection{smale1965,
  title     = {Diffeomorphisms with Many Periodic Points},
  author    = {Smale, S.},
  year      = 1965,
  booktitle = {Differential and Combinatorial Topology},
  publisher = {Princeton University Press},
  address   = {Princeton},
  pages     = {63--80}
}

@book{palisTakens1993,
  title     = {Hyperbolicity and Sensitive Chaotic Dynamics at Homoclinic Bifurcations},
  author    = {Palis, J. and Takens, F.},
  year      = 1993,
  publisher = {Cambridge University Press},
  address   = {Cambridge}
}

@book{katokHasselblatt1995,
  title     = {Introduction to the Modern Theory of Dynamical Systems},
  author    = {Katok, A. and Hasselblatt, B.},
  year      = 1995,
  publisher = {Cambridge University Press},
  address   = {Cambridge}
}

@article{wilczakZgliczynski2003,
  title   = {Heteroclinic Connections between Periodic Orbits in Planar Restricted Circular Three-Body Problem---A Computer Assisted Proof},
  author  = {Wilczak, D. and Zgliczy{\'n}ski, P.},
  year    = 2003,
  journal = {Commun. Math. Phys.},
  volume  = {234},
  pages   = {37--75}
}

@article{zgliczynskiGidea2004,
  title   = {Covering Relations for Multidimensional Dynamical Systems},
  author  = {Zgliczy{\'n}ski, P. and Gidea, M.},
  year    = 2004,
  journal = {J. Differ. Equ.},
  volume  = {202},
  pages   = {32--58}
}

@article{johansson2017arb,
  title   = {Arb: Efficient Arbitrary-Precision Midpoint-Radius Interval Arithmetic},
  author  = {Johansson, F.},
  year    = 2017,
  journal = {IEEE Trans. Comput.},
  volume  = {66},
  number  = {8},
  pages   = {1281--1292}
}

@misc{pythonflint,
  author       = {{python-flint developers}},
  year         = {2026},
  title        = {python-flint: Python Bindings for {FLINT} and {Arb}},
  howpublished = {\url{https://github.com/flintlib/python-flint}}
}

@article{anifrani1995universal,
  title   = {Universal Log-Periodic Correction to Renormalization Group Scaling for Rupture Stress Prediction from Acoustic Emissions},
  author  = {Anifrani, J.-C. and Le Floc'h, C. and Sornette, D. and Souillard, B.},
  journal = {J. Phys. I (France)},
  volume  = {5},
  number  = {6},
  pages   = {631--638},
  year    = {1995},
  doi     = {10.1051/jp1:1995156}
}

@article{mcdonaldFractalBasinBoundaries1985,
  title   = {Fractal basin boundaries},
  author  = {McDonald, Steven W. and Grebogi, Celso and Ott, Edward and Yorke, James A.},
  journal = {Physica D},
  volume  = {17},
  pages   = {125--153},
  year    = {1985},
  doi     = {10.1016/0167-2789(85)90001-6}
}

@article{burkinMatryoshkaMultistableSystem2026,
  title   = {Matryoshka multistable system with uncountable infinity of attractors and closed-form nonlinearity},
  author  = {Burkin, Igor and Rybin, Vyacheslav and Babkin, Ivan and Kuznetsova, Oksana and Butusov, Denis and Karimov, Artur},
  journal = {Chaos Solitons Fractals},
  volume  = {212},
  pages   = {119000},
  year    = {2026},
  doi     = {10.1016/j.chaos.2026.119000}
}

@article{escalanteGonzalezMultistableSystemsHidden2020,
  title   = {Multistable Systems with Hidden and Self-Excited Scroll Attractors Generated via Piecewise Linear Systems},
  author  = {Escalante-Gonz{\'a}lez, R. J. and Campos, E.},
  journal = {Complexity},
  volume  = {2020},
  pages   = {7832489},
  year    = {2020},
  note    = {DOI to be checked}
}

@article{escalanteGonzalezMultistableSystemsNested2022,
  title   = {Multistable Systems with Nested Hidden and Self-Excited Double Scroll Attractors},
  author  = {Escalante-Gonz{\'a}lez, R. J. and Campos, E.},
  journal = {Eur. Phys. J. Spec. Top.},
  volume  = {231},
  pages   = {351--357},
  year    = {2022},
  note    = {DOI to be checked}
}

@article{desouzaBasinsAttractionChanges2005,
  title   = {Basins of Attraction Changes by Amplitude Constraining of Oscillators with Limited Power Supply},
  author  = {de Souza, S. L. T. and Caldas, I. L. and Viana, R. L. and Balthazar, J. M. and Brasil, R. M. F.},
  year    = 2005,
  month   = nov,
  journal = {Chaos, Solitons \& Fractals},
  volume  = {26},
  number  = {4},
  pages   = {1211--1220}
}

@article{defreitasMultistabilityBasinBoundary2004,
  title   = {Multistability, Basin Boundary Structure, and Chaotic Behavior in a Suspension Bridge Model},
  author  = {de Freitas, M. S. T. and Viana, R. L. and Grebogi, C.},
  year    = 2004,
  journal = {International Journal of Bifurcation and Chaos},
  volume  = {14},
  number  = {3},
  pages   = {927--950},
  doi     = {10.1142/S0218127404009636}
}

@article{desouzaNoiseinducedBasinHopping2005,
  title   = {Noise-Induced Basin Hopping in a Gearbox Model},
  author  = {de Souza, S. L. T. and Caldas, I. L. and Viana, R. L. and Batista, A. M. and Kapitaniak, T.},
  year    = 2005,
  journal = {Chaos, Solitons \& Fractals},
  volume  = {26},
  number  = {5},
  pages   = {1523--1531}
}

\end{document}